\documentclass[preprint]{aastex}
\pdfoutput=1
\begin{document} 

\title{{\it Hubble Space Telescope} and Ground-based Observations of V455 
Andromedae Post Outburst\altaffilmark{1}}

\author{Paula Szkody\altaffilmark{2}, 
Anjum S. Mukadam\altaffilmark{2},
Boris T. G\"ansicke\altaffilmark{3},
Arne Henden\altaffilmark{4},
Edward M. Sion\altaffilmark{5},
Dean M. Townsley\altaffilmark{6},
Damian Christian\altaffilmark{7},
Ross E. Falcon\altaffilmark{8},
Stylianos Pyrzas\altaffilmark{9},
Justin Brown\altaffilmark{2},
Kelsey Funkhouser\altaffilmark{2}}

\altaffiltext{1}{Based on observations
made with the NASA/ESA {\it Hubble Space
Telescope}, obtained at the Space Telescope Science Institute, which is
operated by the Association of Universities for Research in Astronomy, Inc.,
(AURA)
under NASA contract NAS 5-26555, with the Apache Point
  Observatory (APO) 3.5-meter telescope, which is owned and operated
  by the Astrophysical Research Consortium (ARC), and the McDonald
  Observatory 2.1m telescope which is owned and operated by the
  University of Texas at Austin.}

\altaffiltext{2}{Department of Astronomy, University of Washington,
  Box 351580, Seattle, WA 98195; 
szkody@astro.washington.edu, anjum@astro.washington.edu}
\altaffiltext{3}{Department of Physics, University of Warwick, Coventry
CV4 7AL, UK; boris.gaensicke@warwick.ac.uk}
\altaffiltext{4}{AAVSO, 49 Bay State Road, Cambridge, MA 02138; arne@aavso.org}
\altaffiltext{5}{Department of Astronomy \& Astrophysics, Villanova University,
Villanova, PA 19085; edward.sion@villanova.edu}
\altaffiltext{6}{Department of Physics \& Astronomy, University of Alabama,
Tuscaloosa, AL 35487; Dean.M.Townsley@ua.edu}
\altaffiltext{7}{Department of Physics \& Astronomy, California State University, Northridge, CA 91330; damian.christian@csun.edu} 
\altaffiltext{8}{Department of Astronomy, University of Texas, Austin,
  TX 78712; cylver@astro.as.utexas.edu}
\altaffiltext{9}{Instituto de Astronomia, Universidad Catolica del Norte, Avenida Angamos 0619, Antofagasta, Chile; stylianos.pyrzas@gmail.com}

\begin{abstract}
{\it Hubble Space Telescope} spectra obtained in 2010 and 2011, three and four 
years after
the large amplitude dwarf nova outburst of V455 And, were combined with
optical photometry and spectra to study the cooling of the white dwarf,
its spin, and possible pulsation periods after the outburst. The
modeling of the ultraviolet (UV)
 spectra show that the white dwarf temperature remains
$\sim$600 K hotter than its quiescent value at three years post outburst, 
and still a few hundred
degrees hotter at four years post outburst. The white dwarf spin at 67.6 s 
and its second harmonic at 33.8 s
are visible in the optical within a month of outburst and are obvious in the
later UV observations in the shortest wavelength continuum and the UV 
emission lines, indicating an origin in high temperature regions near the 
accretion curtains. The UV light curves folded on the spin period show a
double-humped modulation consistent with two-pole accretion. 
The optical photometry two years after outburst shows a group of frequencies 
present at shorter
periods (250-263 s) than the periods ascribed to pulsation at quiescence,
and these gradually shift toward the quiescent frequencies (300-360 s) as time 
progresses past outburst. The most surprising result is that 
the frequencies near this period 
in the UV data are only prominent in the emission lines, not the
UV continuum, implying an origin away from the white dwarf photosphere.   
Thus, the connection of this group of periods with non-radial pulsations
of the white dwarf remains elusive.
\end{abstract}

\keywords{stars:binaries: close --- stars: individual: HS 2331+3905,
  V455 And --- stars: novae, cataclysmic variables}

\section{Introduction} 

V455 And is a unique cataclysmic variable that was first 
discovered in the Hamburg Quasar
Survey (HS 2331+3905; Hagen et al. 1995). During followup observations
since that time, it was discovered that V455 And contains one of the coolest 
white 
dwarfs among all cataclysmic variables, implying a very low secular mean 
accretion rate (Townsley \& G\"ansicke 2009). The data also identified
six different periodicities 
(Araujo-Betancor et al. 2005 (hereafter AB05), G\"ansicke 2007, Bloemen et al.
 2013). Partial eclipses in
the optical light curve revealed an inclination near $\sim$75 deg and an 
orbital period of 81.08 min, near the period minimum (G\"ansicke et al. 2009),
while a photometric period at 83.38 min was also evident, ascribed
to superhumps in a precessing disk.
A longer spectroscopic period of 3.5 hr that drifts on timescales of days was 
observed. A stable short period at 67.62 s and its second harmonic ($2f$)
were found from Discrete Fourier Transform (DFTs) of long data
sets, and attributed to the rotation of the white dwarf, thus identifying this
system as an Intermediate Polar 
(IP){\footnote{asd.gsfc.nasa.gov/Koji.Mukai/iphome/catalog/alpha.html}}. 
In addition,  
a beat period from the spin and the spectroscopic period
(at 67.25 s) and its harmonic were visible. 
Lastly, a broad range of periods between
300-360 s was apparent during quiescence and attributed to non-radial
pulsations of the white dwarf. The width of this broad feature could be
due to unresolved multiplets or a lack of coherence.

As a dwarf nova, V455 And was observed to have an
 outburst in 2007 September when it increased in brightness by eight magnitudes
(Samus et al. 2007; Broens et al. 2007). Like 
all short orbital period dwarf novae, the outbursts are infrequent and
of high amplitude (Howell et al. 1995). The short period is likely the reason
why this system is unique among IPs in having such a high amplitude outburst
and a visible white dwarf. This sole outburst provides
an opportunity to study the heating of a white dwarf from the
outburst (Godon et al. 2006, Piro et al. 2005) and its subsequent cooling, 
as well
as the effect of the outburst on the possible non-radial pulsation. Single,
non-accreting white dwarfs take evolutionary timescales (millions of
years) to cool across the instability strip
(Kepler et al. 2005; Mukadam et al. 2013),
whereas dwarf novae outbursts allow a study of cooling on timescales
of a few years (Mukadam et al. 2011b; Szkody et al. 2012).

Since the accretion disk contaminates the optical light of dwarf novae at
quiescence, the ultraviolet is the best wavelength regime to determine
the parameters of the white dwarf. An ultraviolet spectrum of V455 And
obtained during a snapshot program with the Space Telescope Imaging 
Spectrograph (STIS) during quiescence in 2002 (AB05)
was modeled with a white dwarf at a temperature of 10,500$\pm$250K along
with broad emission lines from an accretion disk viewed at high inclination.
A distance of 90$\pm$15 pc was provided by the model fit. Unfortunately,
this spectrum was too short (700 s) to do any analysis for pulsations.

Our {\it Hubble Space Telescope (HST)} and optical monitoring program was 
designed to determine the post-outburst
cooling of the white dwarf as well as to study the effects of
heating during outburst and subsequent cooling on the observed periods.

\section{Observations and Data Reduction}

Two sets of {\it HST} observations (2010 October 14 and 2011 September 25)
with coordinated ground optical observations 
were obtained three and four years after outburst. The details are provided 
below,
and a summary of the observations is presented in Table 1. Photometric data
have also been collected since 2007 that allow for a comparison of the data
acquired in coordination with the {\it HST} observations to that at quiescence (2003) and
further from outburst. These data are summarized in Table 2.

\subsection{{\it HST} Observations}

The 2010 observation was scheduled for five orbits using the Cosmic Origins
Spectrograph (COS), with the first four
using the G160M grating and the last one with the G140L grating. Unfortunately,
the pointing was lost on the middle three orbits so only the first and last datasets
could be used. The G160M spectra cover 1388-1559\AA\ and 1577-1748\AA\
 with a resolution of
about 0.07\AA,
and the G140L spectra cover 1130-2000\AA\ with a resolution of about 0.75\AA. The timespan
of the good data resulted in orbital phase coverage of 0.18-0.68 for G160M and 
0.92-0.51 for G140L, using the ephemeris provided in AB05.
In 2011, there were two COS orbits with G140L, covering orbital phases of
0.46-0.86 and 0.65-0.25.

All data were obtained in time-tag mode and analyzed using PyRAF routines
from the STSDAS package HSTCOS (version 3.14). From trials of various extraction widths to
optimize the signal-to-noise ratio (S/N), the best values of 27 pixels for the 
G160M and 41 pixels for
the G140L were used. Different light curves were created by binning over
1-5 s timescales and summing over different wavelength regions i.e. with and
without emission lines. To search
for periodic variability via DFT analysis, these light curves were changed into 
fractional amplitude by dividing by the mean and then subtracting one.

The best-fit periods were determined by subjecting the fractional intensity
light curve to least-squares fitting.
An empirical method used in past data analysis (Kepler 1993) was employed to find the
3$\sigma$ limit of the noise. This involved subtracting the best-fit periods,
shuffling the residual intensities
to obtain a pure white-noise light curve, using the DFT of this light
curve to obtain an average (1$\sigma$) amplitude, and then 
repeating this ten times to
derive the 3$\sigma$ value. Computing the 3$\sigma$ white noise limit 
enables confidence that
the signal was not randomly generated, and determines 
which peaks in the DFT can be
safely ignored.

\subsection{Optical Data}

Ground-based optical telescope time  was coordinated with the HST observations. The American
Association of Variable Star Observers (AAVSO)
monitored the brightness for weeks preceding both HST scheduled
times (these data can be viewed from their archive 
site{\footnote{http://www.aavso.org}}). In 2010, three nights of observations with
the Las Cumbres Observatory Global Telescope network 2m Faulkes Telescope North
(FTN) were obtained on October 12, 14 and 16. The Merope CCD was used with a SDSS g
filter and 10 s integrations on October 12, while the Spectral CCD was used on
October 14 and 16. The 2.1m telescope at McDonald Observatory (McD) obtained data on
October 14, 15 and 18 using the Argos CCD camera (Nather \& Mukadam 2004) 
with a BG40 filter and 5-10 s integrations.
The 3.5m telescope at Apache Point Observatory (APO) obtained a 10 minute spectrum
on October 2 using the Double Imaging Spectrograph (DIS) in low resolution mode
(coverage of 3800-9000\AA\ at a resolution of about 4.8\AA\ in the blue and
about 9\AA\ in the red).

In 2011, the Agile CCD camera (Mukadam et al. 2011a) 
with a BG40 filter was used at the  APO 3.5m 
with 5 s integrations
 on September 25 and 26 to obtain light curves. These data were presented
in Silvestri et al. (2012) but were inadvertently labelled with the year 2010 instead
of 2011.

In addition to the ground observations coincident with the HST observations,
long term photometric monitoring was also accomplished with these facilities as well
as the University of Washington Manastash Ridge Observatory (MRO) 0.76m 
telescope
using a 1024x1024 SITe CCD with a BG40 filter, and the 1.2m Kryoneri telescope in
Korinth, Greece using a 516x516 Photometrics CCD with either a V filter or no 
filter (Table 2).

The CCD data were analyzed using IRAF{\footnote{IRAF is distributed by the National 
Optical Astronomy Observatory,
which is operated by the Association of Universities for Research in Astronomy, Inc.,
under cooperative agreement with the National Science Foundation.}} routines to
flat field and bias correct the images and obtain sky-subtracted count rates
for V455 And and comparison stars on the same frames. The mid integration times
were extracted from the headers and converted to Barycentric Dynamic Time (TDB).
The light curves that were then created were treated in the same manner as the 
HST light curves,
with a conversion to fractional intensity for computing DFTs, and subsequent
least squares analysis to determine the best-fit periods and establish the
3$\sigma$ noise limit. 

\section{Results}

The spectra were analyzed for temperature changes in the white dwarf as well as for 
system periodicities in comparison to the parameters of V455 And at quiescence.

\subsection{White Dwarf Temperatures}

The COS G140L spectra obtained at three and four years after the dwarf nova outburst
exhibit the sharp upturn in flux at long wavelengths and the quasi-molecular 
H$_{2}$ absorption at 1600\AA\
 that are typical of a cool white dwarf. These spectra are
plotted along with the STIS spectrum obtained at quiescence (AB05) 
in Figure 1. The increased continuum even at four years after outburst is
evident, along with strong emission lines of 
CIV,CIII,CII (1550,1175,1335\AA), SiIV,SiIII (1400,1300\AA), NV (1240\AA), and 
HeII (1640\AA). Using the same model spectra that were calculated from Hubeny (1988)
and Hubeny \& Lanz (1995) in AB05, and fixing the same gravity to log g=8.0
and photospheric abundances to 0.1 solar, a temperature of 11,100K was determined
for the 2010 data. Figure 1 shows the contributions of 11,100 and 10,600K
white dwarfs. The 2011 spectra fall in-between these values. The error bars on
the fits are on the order of $\pm$250K due to the sensitivity of the
temperature to the quasi-molecular H$_{2}$ absorption at 1600\AA. While
the lack of knowledge of the source and contribution of the FUV emission
impact all the fits, it is clear that the temperature and UV flux are elevated
from the quiescent values. The low temperature of the white dwarf in V455 And
is consistent with that expected for a 0.6M$_{\odot}$ white dwarf at its
orbital period, with a time-averaged accretion rate 
$\sim$3$\times$10$^{-11}$ M$_{\odot}$ yr$^{-1}$ as expected from 
angular momentum losses from gravitational radiation (Figure 3 in 
Townsley \& Bildsten 2003). 

The optical magnitudes provided by the AAVSO near the time of the 2010 observations
were V$\sim$ 16.0 while the quiescent magnitude reported in AB05 is V$\sim$16.4.
The variability of V455 And is on the order of 0.2 mag so the optical brightness
(due to the accretion disk and heated white dwarf) is consistent with the
increased continuum brightness in the UV at three years past outburst. 
The optical spectrum obtained on 2010 October 2 (Figure 2) shows a similar
spectrum to the quiescent one shown in AB05, albeit with slightly increased
continuum, line EWs (Table 3), and FWHM values (1678 and 2037 km s$^{-1}$
for H$\alpha$ and H$\beta$ compared to 1312 and 1475 km s$^{-1}$ in AB05).
The most notable differences are in the stronger blue continuum shortward
of 3800\AA\ and increased H$\alpha$ flux. However, the comparison is not
perfect as the spectrum shown in AB05 is an average of many spectra 
over an orbit while Figure 2 is only one spectrum.

By 2011 September,
the optical brightness ranged from 16.0-16.4, indicating values closer to pre-outburst.
The two datasets indicate that the cooling time for the white dwarf in V455 And 
is greater than four years.

Using the two temperatures from the COS {\it HST} spectra (11,100$\pm$250 K and
10850$\pm$300 K), together with the corresponding days since outburst (1125 and
1471), and the quiescent temperature of $T_{eff,0}=10,500$ K, 
the formulation given in Piro et al. (2005) allows an estimate
of the mass accreted in the outburst. 
Due to the power-law nature of the late-time cooling, as shown in equation 20 of Piro et al. (2005),
$\delta T/T_{eff,0} = (t_{\rm late}/t)^{0.81}$, where 
$\delta T=T_{eff}(t)-T_{eff,0}$, the two
temperature measurements during the cooling provide two estimates of the 
late-time cooling timescale
$t_{\rm late}=33\pm14$ and $22\pm19$~days, for an average 
$t_{\rm late}$ of $27\pm13$~days. This
estimated cooling curve and uncertainty band is shown in Figure 3.
Using this average value along with $\log g=8$ in equation 22 of Piro et al. (2005) yields an accreted
mass of 1.5$\pm$0.6$\times$10$^{-9}$M$_{\odot}$. Due to the uncertainty
of the white dwarf mass, this value likely has an additional uncertainty by a factor
of two. This accreted mass is consistent with the upper limit obtained
from EQ Lyn
after its outburst (Mukadam et al. 2011b).

\subsection{White Dwarf Velocity}

While there are no metal absorption lines evident from the white dwarf in the
UV spectrum (Figure 1),
we attempted to obtain a velocity curve from the strong emission lines that
are present.
All the data from 2010 and 2011 were binned into 0.10 phase bins and
the velocities of CIII, NV, SiIII, CII, SiIV, and CIV lines were measured
using the centroid and Gaussian fitting routines under the {\it splot}
routine in IRAF. The equivalent widths of the lines
are given in Table 3 along with the values at quiescence (from AB05). The
quiescent values were obtained from a single 700 s STIS exposure, while
our COS orbits cover 0.5 of the binary orbital period in 2010 and 0.7 in 2011. 
The
phase coverage shows the large range in variation of the line strengths
during the orbit. However, no consistent velocities could be determined
at comparable phases between the two observation times, nor even at overlapping phases
within the two HST orbits during each observation. This result is similar to
that found from quiescent optical spectra that included orbital coverage, 
where AB05 found the dominant velocity variation in the lines to be near  
3.5 hr. This long period was not
coherent as it showed phase drifts of days. Our short data strings do not
allow us to obtain adequate phase coverage of a 3.5 hr period, but it appears 
the UV lines 
do not originate in a region that is primarily involved in the orbital motion.

\subsection{Periodicities in 2010 and 2011}

The light curves and DFTs created from the COS and optical photometry 
obtained close
in time to the UV data were searched for the spin and pulsation periods.
Figure 4 shows the intensity light curve and DFT from the 2010 October 14
COS data (orbit 1 with G160M and orbit 5 with G140L) with time bins
of 1s and excluding the strong emission lines. Wavelength ranges of
1410-1535, 1579-1632, and 1648-1748\AA\ were used. The prominent spin at
67.6 s and its second harmonic at 33.8 s are easily visible in the light curve
and the DFT. A broad period around 274 s is visible in the expanded
version of the DFT (middle panel) that is above the 3$\sigma$ noise level
of 15 mma. Figure 5 shows the optical data obtained the same night as the COS
data. The 6 hr length of the optical datastream allows a good view of the
superhump variability as well as the spin modulation. These periods show
up easily in the DFT, as well as a prominent broad period at about 280 s.
Figure 6 shows the combined DFT of all the optical data (Table 1) obtained
within a few days of the COS data. The broad pulsation period ranges from
266-295 s over the course of five days.
Due to the similarity of this period to the 274 s one visible 
in the COS data, we
made a first assumption that these periods originate from the same
source. Further evidence for this assumption comes from the
{\it GALEX} NUV (1750-2800\AA) data obtained on 2010 August 26 and September
21 (Silvestri et al. 2012) where periods at 272 s (amplitude of 34 mma) on 
August 26  and 278 s (amplitude of 24 mma) on September 21 were evident.

In efforts to locate the source of the spin and pulsation periods, we
created light curves at a variety of different wavelengths and tried using pure
continuum regions as well as pure emission line regions. Using
the highest resolution G160M data,
we sampled wavelengths shortward and
longward of the upturn in the flux near 1650\AA\ (presumably due to the white
dwarf flux distribution for its 11,000K temperature; Figure 1). The resulting
light curves and DFTs are shown
in Figure 7. The top panels are for 340\AA\ that include lines and continuum
regions over the entire available spectrum. The second panels cover 155\AA\ of
the continuum blueward of the upturn excluding the emission lines. The third 
panels cover
90\AA\ of the longest wavelengths where the white dwarf contributes the most
flux. The bottom panels cover 170\AA\ of the shortest wavelength regions
including both continuum and lines. The FUV continuum below 1600\AA\ is
precisely the region that AB05 could not explain with their model of
a WD + 6500K disk + L2 secondary star. This could be a hot disk, boundary layer,
or magnetic interaction zone. The strongest amplitudes of the spin and its 
harmonic are evident in
the wavelengths shorter than 1620\AA\ and have maximum amplitude
 in the continuum
only panels (second from top). Most surprising, the possible pulsation 
feature (near 0.0035 Hz)
is evident only in the panels that contain emission lines as well as
continua (top and bottom panels), and more importantly, it is absent from the
continuum only panels (second and third panels). These results imply that the origins
of the spin as well as the pulse light are not directly from
the photosphere of the white dwarf. 

We also used the one orbit of G140L grating data to construct a light
curve using only the emission lines (excluding Ly$\alpha$ which is primarily
geocoronal), shown in Figure 8. This plot confirms the result from
the G160M data by showing a much larger amplitude period at 276 s in the
emission lines than in
the continuum plot (Figure 4). We attempted to determine if the lines could
result from the disk reprocessing of the pulsed light from the white dwarf by
comparing the DFTs and phases of the blue versus the red wings of the combined
emission lines. The results yielded a period difference of 20 s. When we force-fit
with a single optimized period, then the phase difference obtained was 50 s. 
This implies much larger distances than
the orbital separation and discounts reprocessing.
However, we do not have sufficient photons in any one
line nor sufficient orbital coverage to obtain a conclusive result. 
In the analysis of optical emission line data obtained only nine months
after outburst, Bloemen et al. (2013) also reported excess power at 288 s in
the periodogram created from the  H$\gamma$ emission line. They
observed this same period
in the continuum between 4520-4670\AA, although at lower amplitude.
However, they reported a different result for the spin amplitudes at
that time, with
a higher amplitude for the second harmonic of the spin in the lines than in the 
continuum.

Folding the white dwarf continuum data in the region of 1420-1520\AA\ 
on the spin period of 67.62 s
and binning to 0.02 phase produces the light curve shown in the top panel
of Figure 9. A clear double-hump modulation is apparent, consistent with
two-pole accretion onto a magnetic white dwarf.

In the following year, on 2011 September 25,
the two orbits of COS data with G140L excluding the emission lines (Figure 10) 
show
only the prominent spin period and its second harmonic, with no obvious presence
of the putative pulsation period in the UV continuum. The optical data obtained
 on the nights of
September 25 and 26 (Figure 11) still show a broad range of periods between 
273-349 s.
The spin amplitudes show no change from the values in 2010.  The stronger
signal from the two orbits with the G140L grating and the closer timing of the 
two {\it HST} orbits
in 2011 compared to the 2010 data (a combination of G160M and G140L
separated by four {\it HST} orbits) should have allowed a better detection of the
pulse period in the continuum. However, past UV data on several accreting 
pulsators at
quiescence (Szkody et al. 2010) have shown that there can be no 
detection in the UV while variability attributed to pulsation is seen in the optical.
While there is a small chance 
that these white dwarfs are exhibiting high $\ell$ g-mode pulsations 
\footnote{Nonradial g-mode pulsations observed in white dwarfs divide the 
stellar surface into zones of higher and lower effective temperature, 
depending on the degree of spherical harmonic $\ell$, thus yielding lower 
optical amplitudes due to a geometric cancelation effect. Increased limb 
darkening at UV wavelengths ensures that modes with 
$\ell\,\leq\,3$ are canceled less effectively, leading to higher 
amplitudes (Robinson et al. 1995).}, the  $\ell $=4 modes do not show a 
significant change in amplitude as a function of wavelength and these modes
have not been unambiguously identified in any of the known ZZ Ceti stars. It is also 
possible that the observed variability may be caused by 
a precessing disk which obscures the view of the
white dwarf. If the pulsation is
somehow related to the interaction of a weak magnetic field with the inner disk, it is of 
note that the Intermediate Polar V842 Cen also shows an optical 
but not a UV period (Sion et al. 2013).

As with the 2010 data, a light curve was constructed using only the 
emission lines (except Ly$\alpha$) from the
two orbits of G140L in 2011 September. The result and the computed DFT is
shown in Figure 12. Comparing this figure with Figure 10 (continuum only for
the same dataset)
shows similar results as the previous year. The group of periods near 275 s
that could be associated with a pulsation is apparent only in the DFT that
isolated photons from
the emission lines. This further reinforces the idea that the origin of this
period is associated with material away from the white dwarf photosphere. 

The spin and second harmonic are present 
in both the continuum and emission line DFTs but have reduced amplitudes in the
 emission line DFT versus the continuum DFT. The bottom panel of Figure 9
shows the 2011 continuum (1420-1520\AA) data folded on the spin period, 
revealing the same double-humped modulation as in the 2010 data. The topmost left
panel of Figure 13 shows the average line profile of the CIV emission feature,
while the lower left panels show the result of folding the data on the spin
period and then binning into 4 bins based on spin phase. All the left panels
show a running average (solid line) with reduced uncertainties computed over
a box length of 25 points, corresponding to a wavelength bin of 2\,$\AA$.
Following the analysis of Bloemen
et al. (2013), these spectra were then divided by the running average of the average
spectrum shown in the top left panel. The resulting flux ratios are plotted in the
right hand panels of Figure 13. In both panels, a component shifting from red to blue is evident.
As was the case for the H$\gamma$ line, the modulations occur at high velocities
(about 1750 km s$^{-1}$), which are about twice the velocity expected at the surface
of a 0.6 M$_{\odot}$ white dwarf rotating at the spin period of 67.62 s. Both the
continuum and line changes are consistent with the two-pole accretion scenario.  

The location of UV emission lines in various types of cataclysmic variables
has been studied with mixed results.  Observations of eclipsing dwarf novae at
high inclination (Szkody, 1987; Mauche et al. 1994) have shown that the
CIV line is generally not affected by the eclipse and hence originates
 from a large volume rather than close to the white dwarf. Studies of
IPs like EX Hya and FO Aqr (Mauche 1999; de Martino
et al. 1999) have shown multiple emission regions. FO Aqr shows similarities
to V455 And, with spin modulations in both the lines and continuua but
with amplitudes that vary with wavelength. De Martino et al. (1999) 
find both a hot component
($\sim$37,000 K) associated with the inner regions of the accretion curtain
or the heated polar regions of the white dwarf and a cooler 12,000 K
component in the outer portions of the curtain
 that can extend out to 6R$_{wd}$. They also
found changes in the spin modulation amplitudes over several years that
they ascribed to changes in the size of the accretion curtain and azimuthal
structure of the disk. Given that V455 And has a relatively high inclination
so that the structure of the disk is important, and it 
is likely undergoing changes
in the accretion curtains due to the outburst, it is difficult to pin
down a particular model.
 
Using the data from October 10, the ratio of UV/optical amplitudes
of the putative pulse period is about 3, similar to the ratio of 2.3 from
 the {\it GALEX} NUV and optical data (Silvestri et al. 2012).            
This ratio is also similar to that from COS and optical data on GW  Lib at three
 years 
past its outburst
(Szkody et al. 2012). However, in GW Lib, the ratio increased to 5 at four years
past outburst, while in V455 And, it appears to decrease between years 3-4.
The ratios after outburst all appear to be less than the typical ratios of 
10-16 seen at quiescence 
(Szkody et al. 2002, 2007) in GW Lib and other accreting pulsating white
dwarfs. Should the post-outburst variability in V455 And be due to non-radial 
pulsations, then the different UV/optical amplitude 
ratios could be understood as an indication of exciting different eigenmodes 
with different indices (see Robinson et al. 1995).

The UV/optical amplitude ratio for the spin period is 9 during both
sets of observations of V455 And. This is
a much larger value than the ratio of 2 found from the longer wavelength NUV 
data obtained in 2010 Aug and Sept. The difference is likely  
due to the greater S/N of
the COS versus {\it GALEX} data; the COS data allows better time
resolution to detect and resolve the spin periodicities. The large ratio of amplitudes 
argues for a location of
the spin component close to, but hotter than a 11,000 K white dwarf. 
This ratio must be used with some caution as the longer optical datasets
allow for a resolution of the spin from the beat period, which can decrease
its amplitude compared to the combined periods in the {$\it HST$} data. A ratio
of 9 would be consistent with a white dwarf model of about 14,000 K (Szkody
et al. 2010) and would also be consistent with the 
optical result of Bloemen et al. (2013), who
concluded that the spin likely originates from the accretion curtains near
the white dwarf, although they also could not produce any detailed model.
This temperature is also in the range found by de Martino et al. (1999) for
the accretion curtains of FO Aqr.

\subsection{Long Term Trends}

Figure 14 shows a compilation of DFTs using optical data obtained 
from one month 
(2007 October) to five years (2012 October) past outburst as well as the
DFT from quiescent data several years before the outburst (Pyrzas, S. et
al. in prep). 
It is clear that the
spin is weakly present even while the disk is still dominating the light at
one month after the outburst, and then its amplitude increases dramatically 
by two years
past outburst when the disk is close to its quiescent level. The
period of the spin is not noticeably altered by the outburst, but the amplitude
ratio of the spin to its second harmonic reverses in the data soon after outburst.

In contrast, the broad range of periods ascribed to pulsations at
quiescence disappears close to outburst and begins to
re-appear by 2009 October, but at a shorter period range than apparent at
quiescence. The observed periods drift to longer timescales as the time from
outburst increases, approaching, but not yet identical to, the quiescent periods
even at five years post outburst. This behavior is not unexpected if the origin
is non-radial pulsations, as a change
from short to long periods is consistent with excitation of different
modes with rapid cooling of the outer envelope. Which pulsation modes are 
excited
depends on the thermal timescale at the base of the convection zone
(Brickhill 1992, Goldreich \& Wu, 1999, Wu 2001, Montgomery 2005), so
as the star cools, the base moves deeper into the white dwarf. However,
this slow drift in pulsation period at the rate of 
d$\nu$/dt\,$\sim$\,-$10^{-12}$\,Hz/s (Townsley et al. 2004) from continued 
cooling of the outer envelope after outburst, is not evident in all the 
accreting white dwarfs that have
been observed so far. EQ Lyn returned to exactly its
pre-outburst pulsation spectrum within 3.3 years following its outburst
(Mukadam et al. 2011b). GW Lib showed two completely different periods
in the five years after its outburst (Szkody et al. 2012; Chote \& 
Sullivan 2013) 
and still has not returned to its
quiescent pulsation modes. V455 And has the coolest white dwarf among all
the accreting pulsators and is one of the few that exists at a quiescent 
temperature
that would place it in the 
instability strip for non-interacting DAV pulsators (Gianninas et al. 2005).
Both EQ Lyn and GW Lib are much hotter at 15,000 K (Szkody et al. 2002,
Mukadam et al. 2013). Arras et al (2006) have suggested that these hotter
systems could have increased He abundance and hence be unstable due to
HeII ionization. This difference in composition could lead to the
different behaviors observed. Alternatively, the very fast rotation period of
V455 And compared to the other accreting white dwarfs could influence its
behavior. Townsley (2010) has shown that if the spin frequency is higher
than the pulsation mode frequency, the Coriolis force alters the observed
modes. 

The major unsolved puzzle in ascribing this period to a non-radial pulsation
of the white dwarf is why the
spin period is so visible in the UV whereas the pulsation period is not,
if they both originate from the white dwarf. The presence of the spin in
the shortest wavelengths and in the emission lines argues for a location
in the accretion curtains. If the $\sim$300 s period originates in these
curtains or in the accretion disk itself, it is not clear how 
a period of this timescale could be sustained for years and how it would 
gradually
shift from shorter to longer periods from two to five years after outburst.

Ortega-Rodriguez \& Wagoner (2007) explored nonradial g-mode disk oscillation modes in
cataclysmic variables, however their model was for optically thick, steady-state
disks and the resulting oscillations are short (tens of seconds). 
Yamasaki et al. (1995) do produce periods
in the range of 70--600\,s with axially symmetric radial p-mode pulsations
trapped in the outer part of an accretion disk, but this area could not
explain the periodicity in the high excitation UV emission lines of V455 And. 
There is a long history of observed 
periodicities in dwarf novae disks but none match the timescales of V455 And.
Warner (2004) provides a review of these periods, which are termed
dwarf nova oscillations (DNOs), longer period DNOs (termed lpDNOs), and 
quasi-periodic oscillations (QPOs). He argues that all of these are
related to magnetic accretion. The DNOs could be caused by magnetic coupling
to the equatorial accretion belt which is rotating at Keplerian velocities
during a dwarf nova outburst. The lpDNOs appear to be related to the rotation of
the white dwarf, where P$_{lpDNO}\sim$1/2P$_{wd}$ and $\sim$ 4$\times$P$_{DNO}$.
The QPOs can be explained by reprocessing or obscuration of the light from the
hot inner regions by a slow moving prograde traveling wave in the inner disk
that creates a vertical thickness of the disk.
Theoretical support for such a wave was provided by Lubow \& Pringle (1993),
and Warner \& Woudt (2002) postulate that this wave could be excited by
the magnetic field interaction with the disk. The 3-D models of Romanova
et al. (2003) of magnetic accretion onto an inclined rotating dipole have
produced QPOs. While this magnetic model seems appropriate since V455 And
shows the stable spin period (and its second harmonic) indicative of two pole
accretion, the timescales do not appear to work out. In most dwarf novae,
the DNOs are on the order of 10 s, the lpDNOs around 40 s, and the QPOs
about 160 s. The Romanova et al. models produce QPOs that are $\leq$P$_{wd}$.
If there is no spun up equatorial belt from the disk that would
produce a short period DNO in V455 And, and we use the spin period
of 67.62 s in equation 23 of Warner \& Woudt (2002) where
P$_{QPO}$/P$_{wd}$ = 10-19, we obtain QPO periods of 676-1284 s, far longer
than those observed. If the timescales can be reproduced correctly, the
magnetic model offers advantages, as the period changes could be related
to the changing radius of the inner disk and the changing vertical height
of a traveling wave could account for some obscuration that affects the
QPO period but not the visibility of the accretion curtains where the
spin period is observed. 
 
\section{Conclusions}

Our UV and optical data on V455 And obtained following its 2007 dwarf
nova outburst have provided insights as well as dilemmas. The results
can be summarized as follows.

\begin{itemize}

\item $\it HST$ spectra at three and four years past outburst show the white
dwarf remained heated by several hundred degrees. Thus, the cooling time for
large amplitude outbursts is greater than four years. The optical
spectra and photometry also showed increased fluxes over quiescent values 
at three years past
outburst but quiescent levels were reached by the fourth year. The cooling curve
obtained from the white dwarf temperatures imply a mass 
$\sim$1.5$\times$10$^{-9}$M$_{\odot}$ was accreted during the outburst.

\item The UV emission lines show large flux variability but no obvious 
correlation with the orbital period.

\item The spin period and its second harmonic show up within a month of outburst
and are consistently present. The second harmonic has greater amplitude
during the year following outburst. 
The strongest amplitudes of
the spin and its second harmonic occur at wavelengths $<$1620\AA, and these periods
are present in the UV emission lines as well. These results imply an
origin from a hot component, possibly the accretion curtain close to
the white dwarf. The UV continuum data folded on the spin period show
a clear double-humped modulation consistent with two-pole accretion.
The CIV emission line also shows changes in shape when phased
on the spin period.

\item A range of periods that are similar to the 300-360 s observed at
quiescence are apparent in the optical data following outburst, but
gradually shift from 250-263 s when they appear two years after outburst
to 312-350 s by five years after outburst. While a period in this range is 
evident in the UV continuum data at three years past outburst, it is 
absent at four years. Most surprising, this period is prominent in the UV 
emission lines in both years, indicating an origin away from the white
dwarf photosphere.

\item While an association of the spin and 300 s periods with the
accretion curtain appears plausible, it is not clear why the spin is
clearly visible in the UV continuum whereas the putative pulsation is not.
If this period originates in the hotter portions of the accretion curtain,
the mechanism for producing a period in this range, sustainable for
years and moving from long to short period as the white dwarf cools,
is not clear.

\end{itemize}

The presence of permanent superhumps
indicates the disk is eccentric, oscillating and precessing. The strong presence of the spin period and its
second harmonic points to an IP with a magnetic white dwarf and
corresponding accretion curtains, as well as disk light,
to take into account. While these components complicate the formulation
of a good system model, the brightness of V455 And means that it can be
easily observed from space and ground to provide the long term datasets
that can ultimately lead to a better understanding of the emission regions
of a magnetic, rapidly rotating, possibly pulsating, accreting white dwarf.

\acknowledgments
We gratefully acknowledge Arlo Landolt and the 
observers of the AAVSO who contributed measurements
during the course of the two sets of HST observations.
This work was supported by NASA grants HST-GO1163.01A, HST-GO11639-01A
and HST-GO12231.01A from the Space Telescope Science Institute, which is
operated by the Association of Universities for Research in Astronomy, Inc.,
for NASA, under contract NAS 5-26555, and by NSF grant AST-1008734.
The research leading do these results has also received funding from the
European Research Council under the European Union's Seventh Framework
Programme (FP/2007-2013)/ERC Grant Agreement n. 267697 (WDTracer).
BTG was supported in part by the UK's Science and Technology Facilities Council
(ST/1001719/1).

\clearpage

\begin{deluxetable}{lclcl}
\tablewidth{0pt}
\tablecaption{Summary of HST and Related Optical Observations}
\tablehead{
\colhead{Date} & \colhead{Obs} & \colhead{Filter} & \colhead{Exp (s)} & 
\colhead{UT time} }
\startdata
2010 Oct 2 & APO & DIS & 600 & 04:48-04:58 \\
2010 Oct 12 & FTN & g & 10 & 08:02:02-12:54:40 \\
2010 Oct 14 & McD & BG40 & 5 & 02:56:59-09:23:39 \\
2010 Oct 14 & FTN & g & 10 & 08:02:48-12:34:47 \\
2010 Oct 14 & HST & G160M & ... & 13:03:22-13:43:22 \\
2010 Oct 14 & HST & G140L & ... & 19:27:22-20:15:22 \\
2010 Oct 15 & McD & BG40 & 10 & 06:47:09-07:50:59 \\
2010 Oct 15 & McD & BG40 & 5 & 07:53:38-09:00:48 \\
2010 Oct 16 & FTN & g & 10 & 06:22:36-07:21:50 \\
2010 Oct 18 & McD & BG40 & 5 & 05:09:16-05:54:36 \\
2011 Sep 25 & APO & BG40 & 5 & 01:48:31-12:17:56  \\
2011 Sep 25 & HST & G140L & ... & 12:33:48-13:05:48 \\
2011 Sep 25 & HST & G140L & ... & 14:09:48-14:57:58 \\
2011 Sep 26 & APO & BG40 & 5 & 01:50:48-07:04:48  \\
\enddata
\end{deluxetable}

\clearpage

\begin{deluxetable}{lclc}
\tablewidth{0pt}
\tablecaption{Summary of Long Term Optical Monitoring Observations}
\tablehead{
\colhead{Date} & \colhead{Obs} & \colhead{Filter} &  
 \colhead{Total Hours} }
\startdata
2003 Aug 14-20 & Kry & clear,V  & 24.8 \\
2007 Oct 17 & APO & BG40 & 1.5 \\
2008 Sep 6-8 & APO,McD & BG40 & 8.3 \\
2009 Oct 7-9 & Kry & clear & 13.0 \\
2010 Sep 7-12 & APO,McD & BG40 & 17.5 \\
2010 Oct 11-18 & FTN,McD & g,BG40 & 17.7 \\
2011 Sep 25-26 & APO & BG40 & 15.7 \\
2012 Jul 21 & APO & BG40 & 3.8 \\
2012 Aug 05 & MRO & BG40 & 4.9 \\
2012 Oct 13 & APO & BG40 & 10.6\\
\enddata
\end{deluxetable}

\clearpage
\begin{deluxetable}{lllc}
\tablewidth{0pt}
\tablecaption{Emission Line Equivelent Widths (\AA)}
\tablehead{
\colhead{Line (\AA)} & \colhead{2010 Oct} & \colhead{2011 Sep} & 
\colhead{2002 Oct\tablenotemark{a}} }
\startdata
CIII (1175) & 81-118 & 77-135 & 111 \\ 
NV (1240) & 30-39 & 27-47 & 42 \\
SiIII (1300) & 23-63 & 16-69 & ... \\
CII (1335) & 52-72 & 48-76 & 67 \\
SiIV (1400) & 63-89 & 60-86 & 52 \\
CIV (1550) & 159-183 & 150-367 & 234 \\
H$\beta$ & 84 & ... & 75 \\
H$\alpha$ & 225 & ... & 185 \\
\enddata
\tablenotetext{a}{values from AB05 at quiescence}
\end{deluxetable}

\clearpage
\begin{figure}
\includegraphics[angle=-90,width=7in]{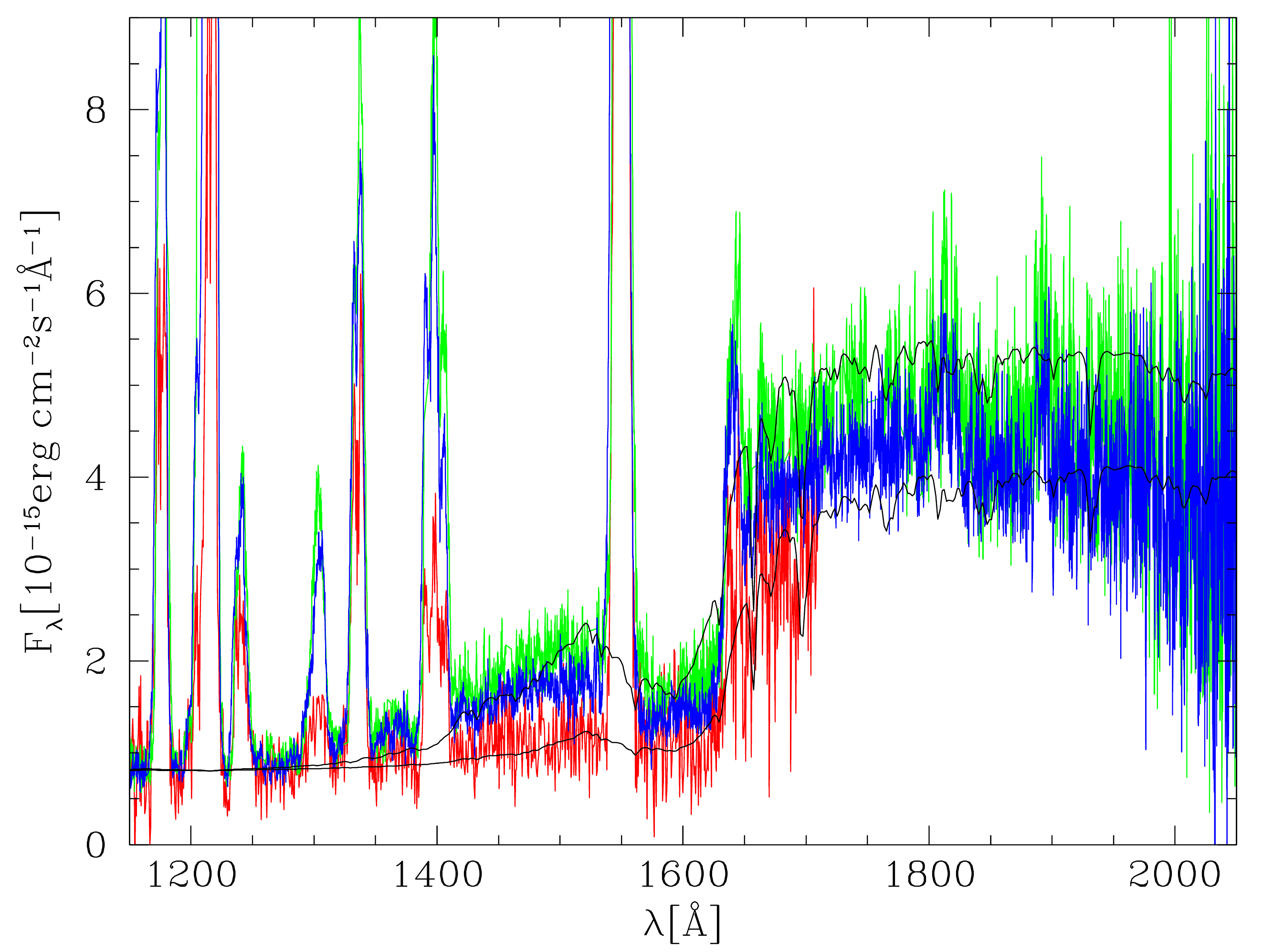}
\caption{COS G140L data from 2010 Oct 14 (top, green), 2011 Sep 25 (middle, blue),
and STIS data from 2002 (bottom, red). Black lines are models for 10,600K (bottom) and
11,100K (top).}
\end{figure}

\clearpage
\begin{figure}
\plotone{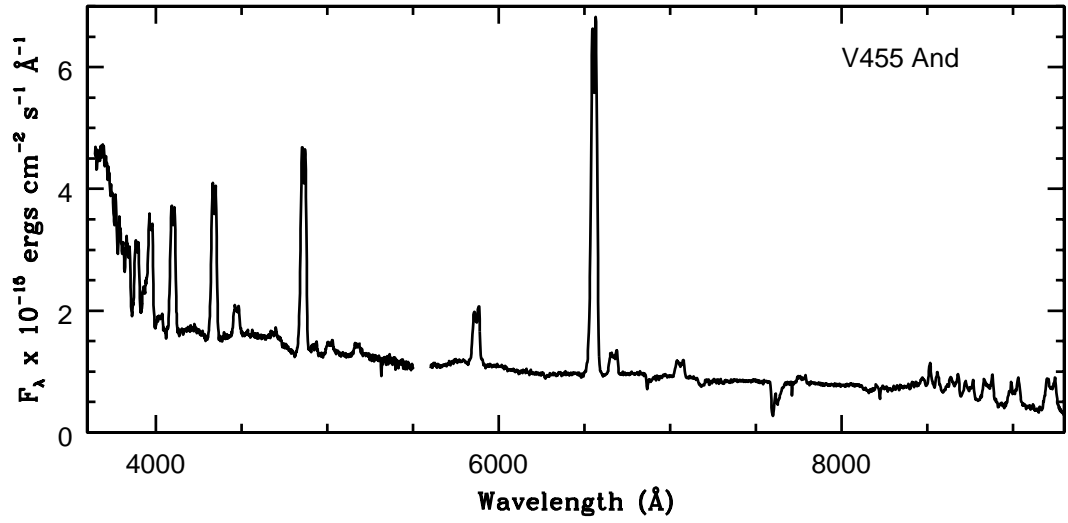}
\caption{APO spectra obtained 2010 Oct 2. Short gap near 5500\AA\ marks
where the dichroic separates the blue and red spectra.}
\end{figure}

\clearpage
\begin{figure}
\epsscale{0.75}
\plotone{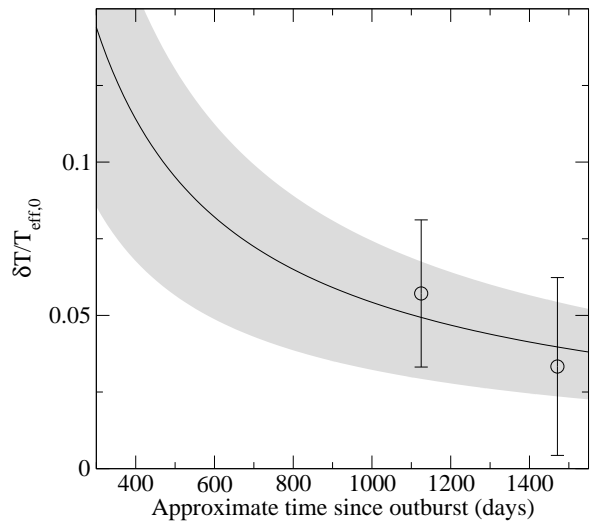}
\caption{Cooling curve based on the two temperature measurements at 3
and 4 years post-outburst and the quiescent temperature $T_{eff,0}$ from 
pre-outburst STIS data.}
\end{figure}

\clearpage
\begin{figure}
\epsscale{0.8}
\plotone{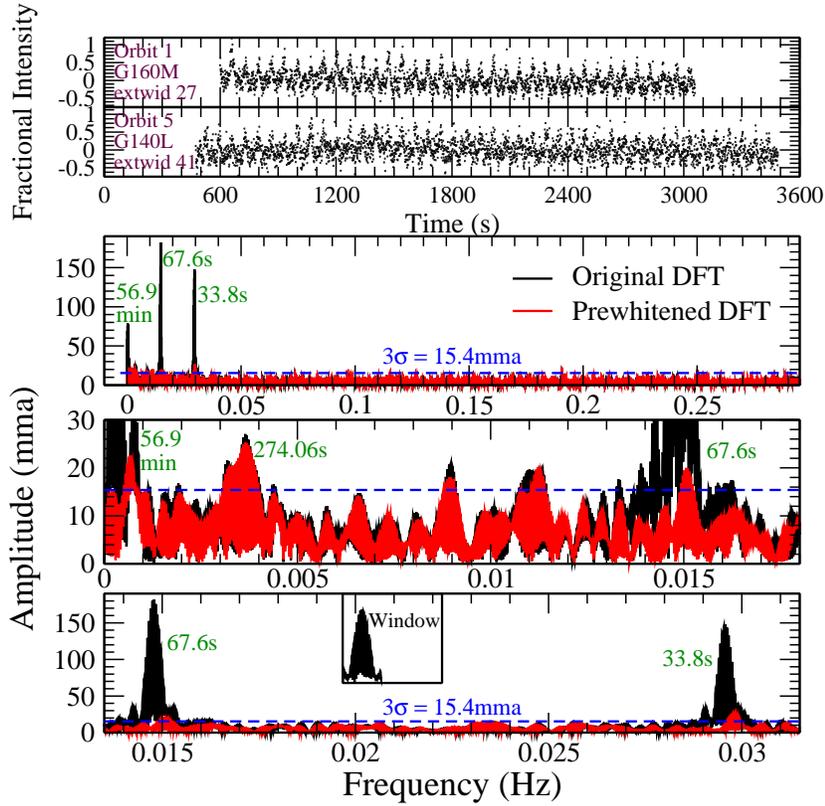}
\caption{COS light curve from 2010 October 14 with 1 s exposures, using
wavelength regions 1410-1535, 1579-1632, and 1648-1749\AA, excluding
emission lines for orbit 1 (G160M) and orbit 5 (G140L). Fractional intensity (top), DFT (middle) with spin period
(67.6 s) and second harmonic (33.8 s) labelled, and expanded high frequency regions
(bottom) with pulsation period (274 s) labelled.}
\end{figure}

\clearpage
\begin{figure}
\epsscale{0.8}
\plotone{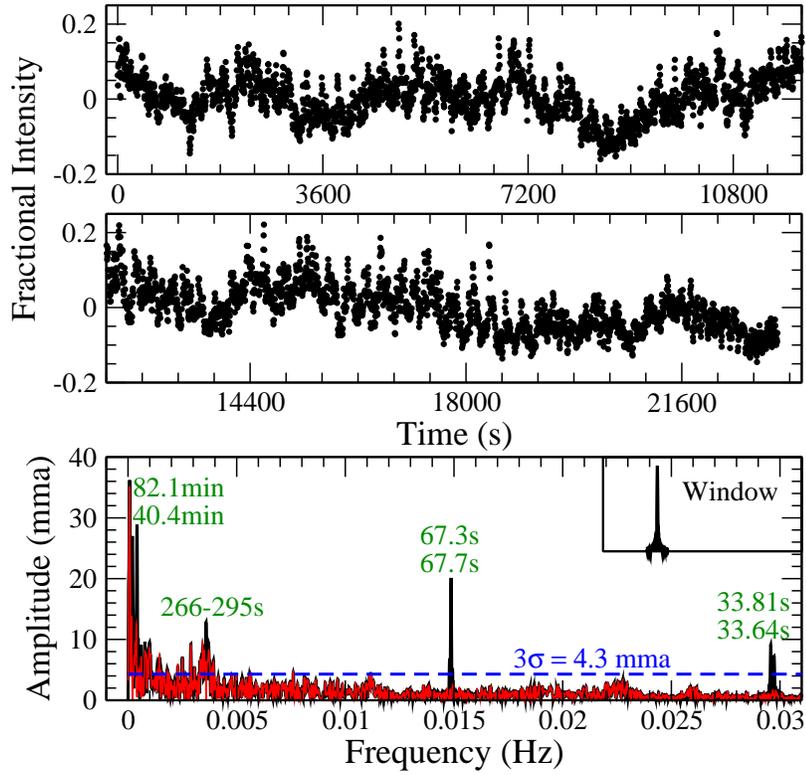}
\caption{Light curve and DFT from McDonald 2010 Oct 14
  with 5 s exposures and BG40 filter. Fractional intensity (top) and
 DFT (bottom).}
\end{figure}

\clearpage
\begin{figure}
\epsscale{0.8}
\plotone{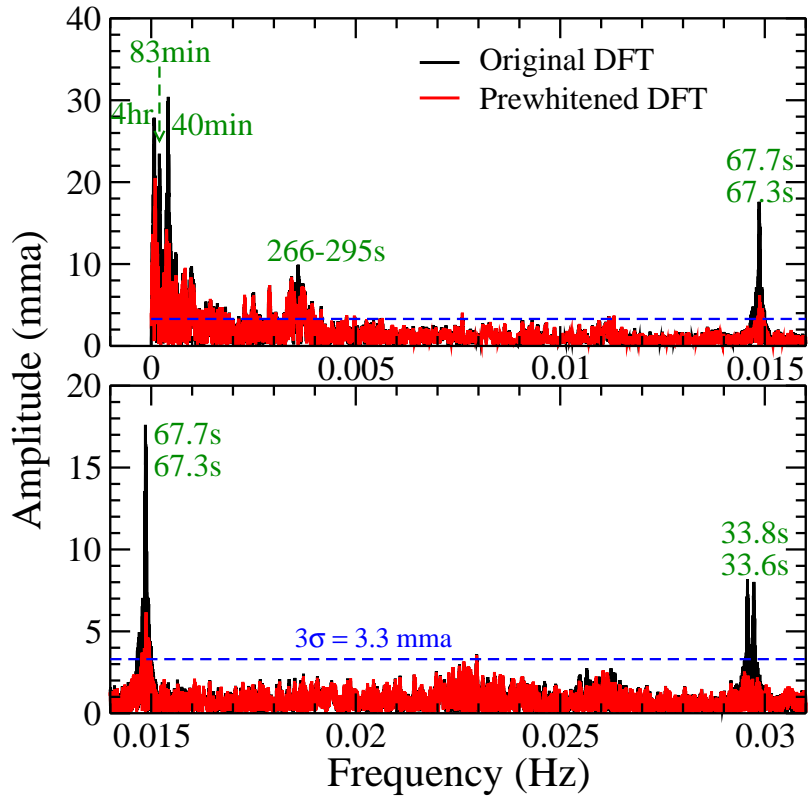}
\caption{Combined DFT from 2010 Oct 12-18 showing the photometric period
at 83 min, the pulse period at 276.5 s, and the spin and beat (67.7, 67.3 s) 
periods and their second harmonics.} 
\end{figure}

\clearpage
\begin{figure}
\plotone{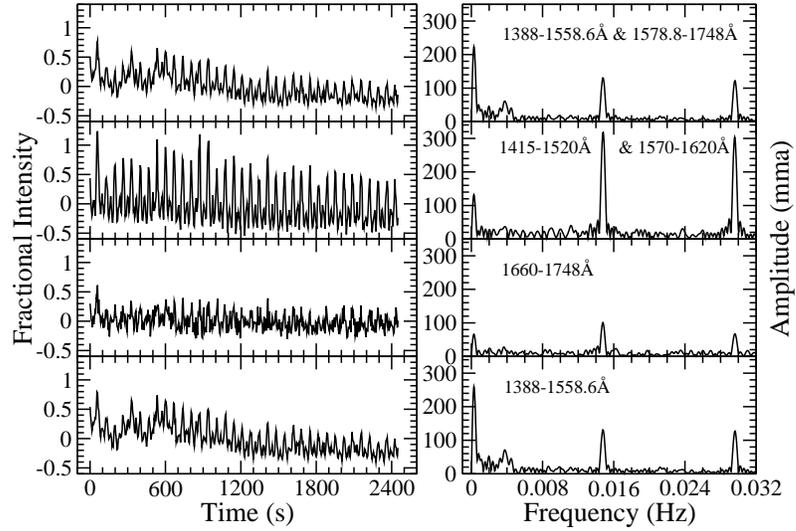}
\caption{Light curves (left) and DFTs (right) of the first COS orbit with
G160M on 2010 Oct 14 constructed with the different wavelength regions as
labelled. Top and bottom panels include continuum and lines (top is entire
range while bottom is restricted to wavelengths where the white dwarf is 
negligible).
Middle
two panels are continuum only, with the third panel being wavelengths where the 
white dwarf dominates and the second where the other component of the FUV 
light dominates.}
\end{figure}

\clearpage
\begin{figure}
\epsscale{0.8}
\plotone{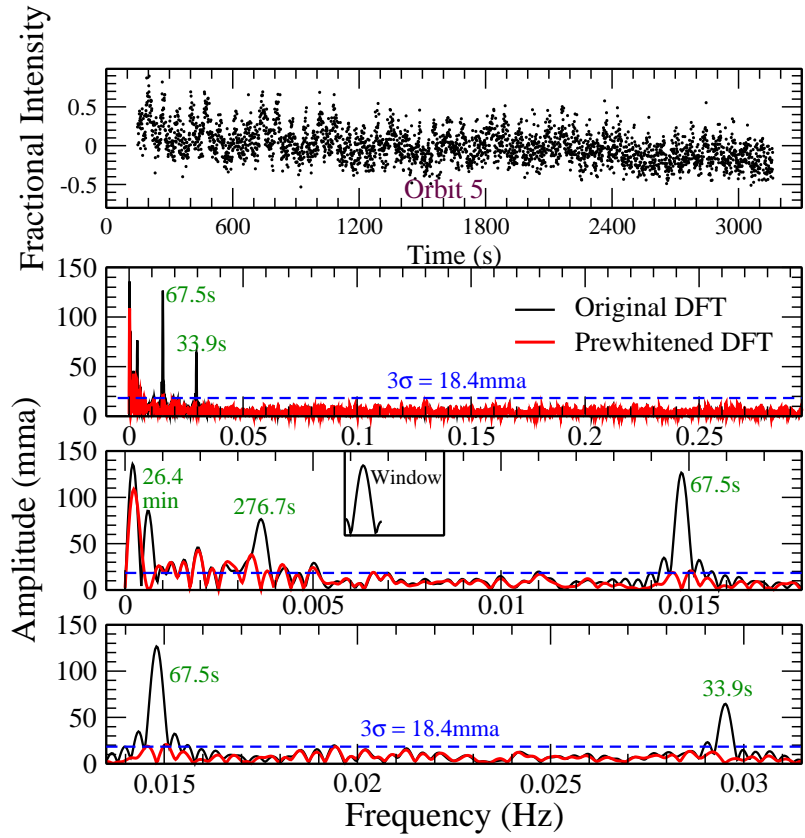}
\caption{COS light curves from 2010 Oct 14 with 1 s exposures, using
only wavelength regions containing emission lines (excluding Ly$\alpha$).
Fractional intensity (top), DFT (middle) with spin period
(67.6 s)
and second harmonic (33.8 sec) labelled, and expanded high frequency regions
(bottom).}
\end{figure}

\clearpage
\begin{figure}
\includegraphics[angle=-90,width=7in]{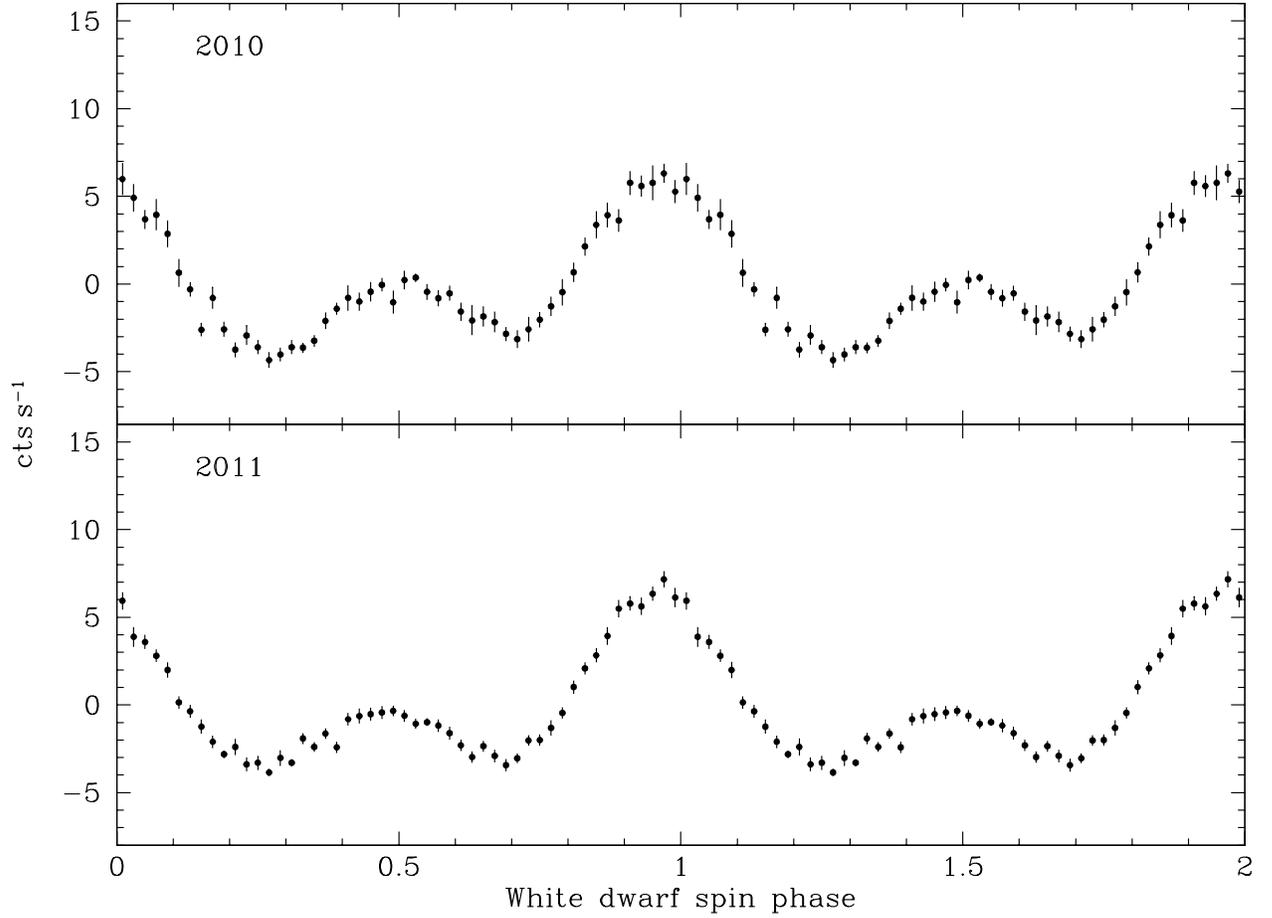}
\caption{COS G140L light curves from UV continuum region 1420-1520\AA\ folded
on the white dwarf spin period of 67.62 s and binned into 0.02 phase bins.
Top panel is  2010 Oct 14 and bottom panel is 2011 Sep 25. Vertical scales
are offset for plotting.}
\end{figure}

\clearpage
\begin{figure}
\epsscale{0.8}
\plotone{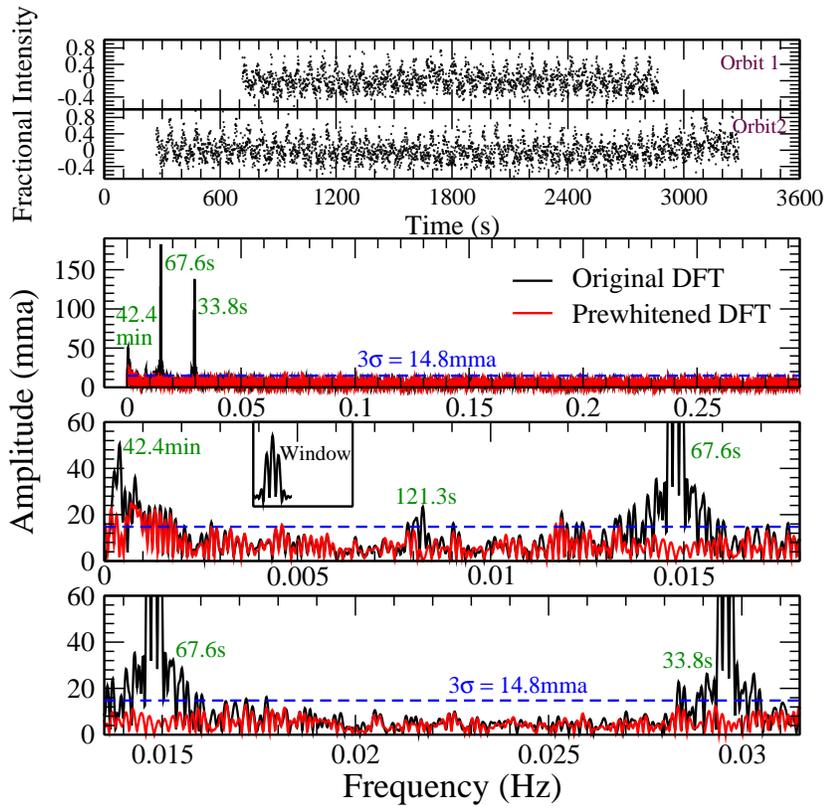}
\caption{COS light curve from 2011 Sep 25 with 1 s exposures, using
wavelength regions 1410-1535, 1579-1632, and 1648-1749\AA, excluding
emission lines. Fractional intensity (top), DFT (middle) with spin period
(67.6 s)
and second harmonic (33.8 sec) labelled, and expanded high frequency regions
(bottom).}
\end{figure}

\clearpage
\begin{figure}
\plotone{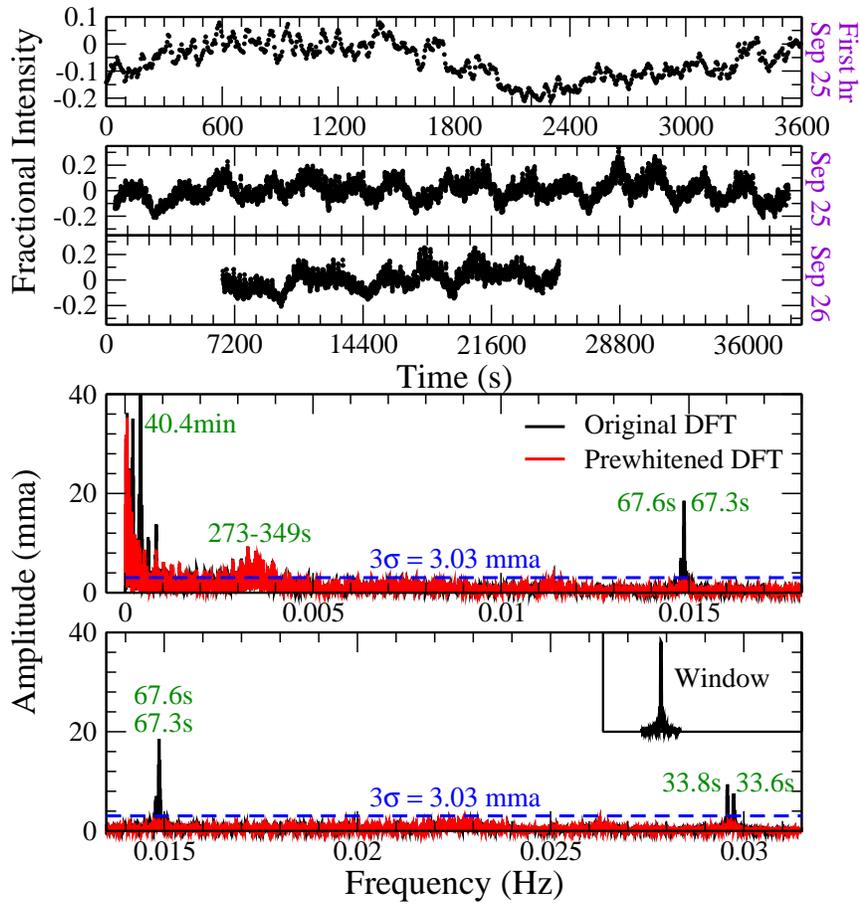}
\caption{Light curves and DFTs from 2011 Sep 25 and 26 APO data. Top 
intensity curve is the expanded first hour of observations to show  
details of the light curve variability.}
\end{figure}

\clearpage
\begin{figure}
\epsscale{0.8}
\plotone{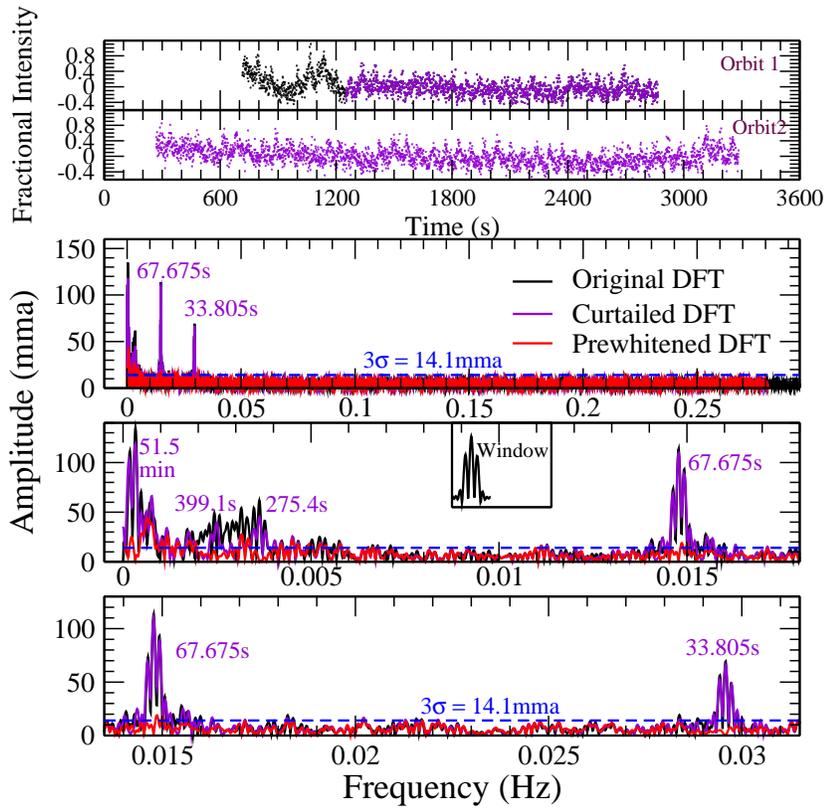}
\caption{COS light curve from 2011 Sep 25 with 1 s exposures, using
only wavelength regions containing emission lines (excluding Ly$\alpha$).
Fractional intensity (top), DFT (middle) with spin period
(67.6 s)
and second harmonic (33.8 s) labelled, and expanded high frequency regions
(bottom). Curtailed DFT is excluding the feature in the first 600 s
of data in Orbit 1.}
\end{figure}

\clearpage
\begin{figure}
\epsscale{1.0}
\plotone{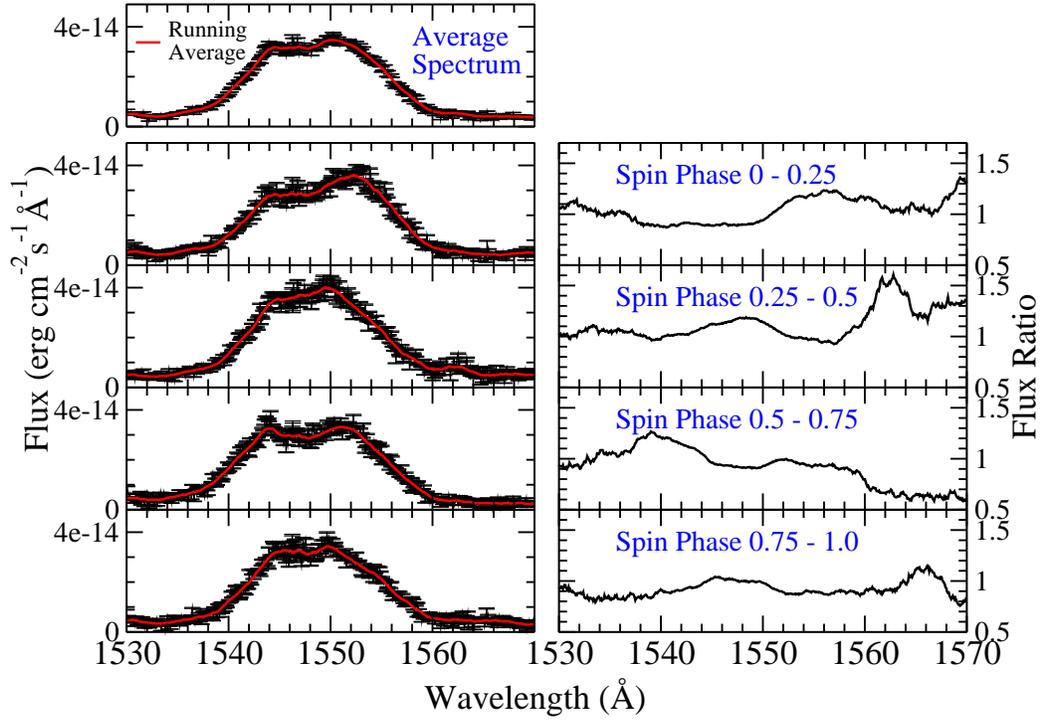}
\caption{2011 Sep 25 COS spectral profile of the CIV line phase-folded and binned
at the spin period of 67.619 s (left panels), with the solid lines indicating a running
average obtained over 2\,$\AA$ (box length 25 points). Right hand panels show these
spin-phased spectra divided by the running average of the average line profile (solid
line in top left panel).   
The lines change from red
peaks (spin phase 0-0.25) to blue (spin phase 0.5-0.75) during the spin cycle.}
\end{figure}

\clearpage
\begin{figure}
\epsscale{1.0}
\plotone{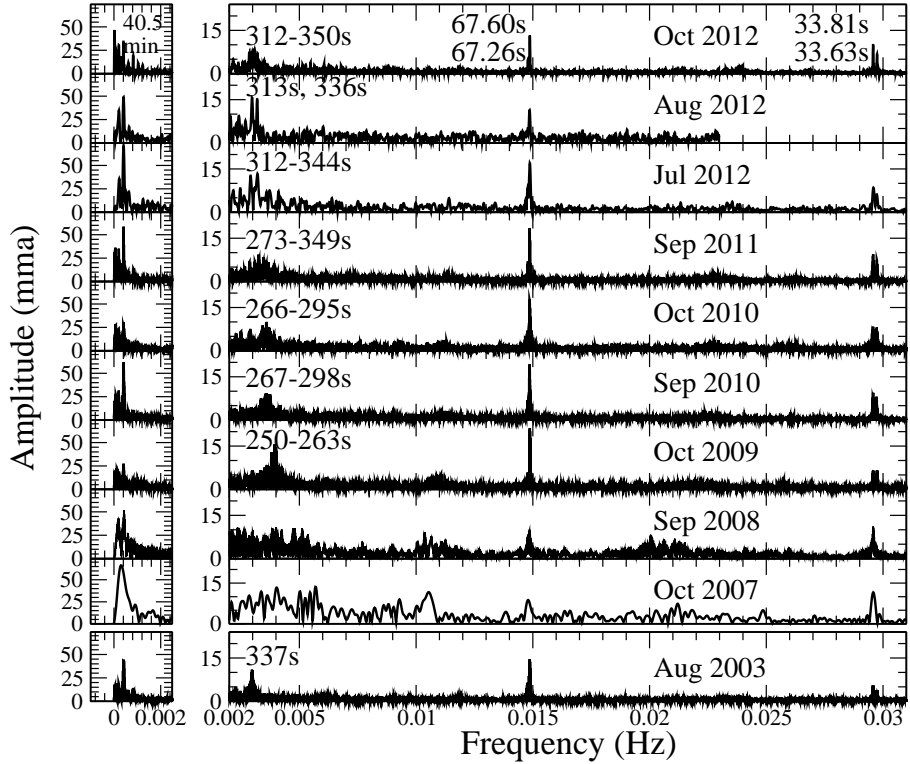}
\caption{Combined DFTs from outburst (Oct 2007) to five years past outburst
(2012) showing the stability of the spin period but the progression of
the $\sim$300 s period from shorter to longer periods as the time from outburst 
increases. The DFT at quiescence is shown at the bottom for comparison. Tables
1 and 2 provide the origin of the data.}
\end{figure}

\end{document}